\documentclass[12pt]{article}

\input epsf
\def\beq{\begin{eqnarray}}
\def\eeq{\end{eqnarray}}

\def\lsim{\mathrel{\rlap{\lower3pt\hbox{\hskip0pt$\sim$}}
    \raise1pt\hbox{$<$}}}         
\def\gsim{\mathrel{\rlap{\lower4pt\hbox{\hskip1pt$\sim$}}
    \raise1pt\hbox{$>$}}}         

\begin{document}


\vskip 1cm
\begin{center}
{\Large \bf  Three-Form Gauging  of axion Symmetries and Gravity }

\vskip 1cm {Gia Dvali\footnote{\it  email:  dvali@physics.nyu.edu}}

\vskip 1cm
{\it Center for Cosmology and Particle Physics, Department of Physics, New York University, New York, NY 10003}\\
\end{center}

\vspace{0.9cm}
Nonlinearly realized Abelian global symmetries can be reformulated as local shift symmetries
gauged by  three-form gauge fields.  The anomalous symmetries of the Standard
Model (such as  Peccei-Quinn or $B+L$) can be dualized to local symmetries
gauged by the Chern-Simons  three-forms of  the Standard Model gauge group.  
In this description the strong CP problem can be reformulated as the problem of a massless
three-form field in QCD, which creates an arbitrary CP-violating constant four-form electric field in the vacuum.  Both the axion as well as the massless quark solutions amount to simply
Higgsing the three-form gauge field, hence screening the electric field in the vacuum.  
This language gives an alternative way for  visualizing the physics of the axion solution
as well as the degree of its vulnerability due to  gravitational corrections.  
Any physics that can jeopardize the axion solution must take the  QCD three-form 
out of the Higgs phase. This can only happen if the physics in question provides an 
additional massless three-form. The  axion then Higgses one combination of the three-forms and the QCD electric field gets partially unscreened, reintroducing the strong CP problem.  Gravity provides such a candidate in form of the Chern-Simons spin connection three-form, which could un-Higgs the QCD three-form in the  absence of additional chiral symmetries. We also discuss  analogous effects for the baryon number symmetry. 

\vspace{0.1in}

\newpage

\section{Introduction}

The global symmetries often play an important role in particle physics models. The well-known examples
are the baryon and the lepton number symmetries, and the Peccei-Quinn (PQ) symmetry \cite{pq}, which playes the crucial role in the dynamical solution of the strong CP problem\cite{theta}. PQ is a nonlinearly realized anomalous symmetry, and consequently implies the existence of a massive pseudo-Goldstone 
particle, an axion $a$ \cite{axion,axion1}.  The strong CP problem is a problem of  vacuum selection,  and can be expressed as the inexplicable smallness  of the CP-violating $\theta$-parameter in QCD Lagrangian.
A non-zero $\theta$ implies a non-zero expectation value of the dual gauge field strength,
$\langle$Tr$F\tilde F\rangle\, \neq 0\, $, which would lead to observable CP 
violation, unless $\theta$ is tiny, $\theta < 10^{-9}$\cite{review}.   The essence of the axion solution to this problem is 
to promote $\theta$  into a dynamical variable by introducing a pseudo-scalar field 
$a$ with the following coupling
\begin{equation}
\label{aff}
a \, {\rm Tr } F\tilde F.
\end{equation}
The axion can emerge as a Goldstone boson as a result of spontaneous PQ symmetry breaking at 
an arbitrarily high scale\cite{invisible,invisible1}.  Alternatively, PQ symmetry may be introduced
as an intrinsically nonlinearly realized symmetry at the field theory level, as it is usually the 
case for string theoretic axions. 

The key point of the PQ axion solution to the strong CP problem is that the minimum of the axion potential is automatically at $\theta\,  =  \,0$,  where the expectation 
value of ${\rm Tr} F\tilde F$ vanishes. This follows from the proof by Vafa and Witten\cite{vw}.

 The standard gauging of the anomalous global symmetries is impossible because of the obvious reasons, but in certain ways they can be physically equivalent  to the gauge symmetries. 
  The  phenomenon investigated in the 
present paper is an alternative gauging of abelian symmetries by an antisymmetric  three-form gauge field $C_{\alpha\beta\gamma}$. 
The best way to understand such a gauging is in the dual language in which the pseudo-scalar 
axion is replaced by an antisymmetric two-form field $B_{\mu\nu}$. In this language the three-form 
$C_{\alpha\beta\gamma}$ gauges the shift symmetry of $B_{\mu\nu}$ and gains a gauge invariant mass by "eating up" the latter.
Thus, by mixing with $B_{\mu\nu}$ (or equivalently, with the axion) the three-form gauge field gets {\it Higgsed}. 

Because the massless three-form carries no propagating degrees of freedom, the number 
of propagating degrees of freedom is both descriptions is one.  Thus,  the net  result of the 
three-form Higgs effect  is that the axion gains a mass (or the potential).  

The PQ symmetry in QCD is  in fact a good example of a shift symmetry  gauged
by a three-form. 
It has been known for a long time\cite{qcdform}  that QCD with no light quarks contains a 
massless three-form field, that can mediate a constant long-range electric field in the vacuum.
The electric field here refers to the non-zero value of the gauge-invariant field strength of $C$, 
which is a four-form. 
The value of this electric field coincides with the expectation value of ${\rm Tr} F\tilde F$. Thus, having a nonzero four-form electric field in the vacuum is completely equivalent to having a non-zero
$\theta$-term, and vice versa.  This fact gives a possibility to reformulate the strong CP 
problem in the three-form language, as the problem of the unnatural smallness of the four-form field
strength in the vacuum.  This formulation gives a simple alternative way for visualizing  
physics of the axion solution and its susceptibility to non-QCD corrections.

   In the presence of axion, the three-form gauges the axion
shift symmetry. Thus, axion solves the strong CP problem by putting the three-form 
into the Higgs phase.  In this phase the four-form electric  field is {\it screened}, and the vacuum is 
automatically CP-conserving.  Thus, the three-form language gives a very simple explanation 
to the fact that the minimum of the axion potential is always at $\theta \, = \, 0$. 
 
 Knowing that PQ symmetry can be understood as intrinsically gauge symmetry, it is natural to ask, 
whether the axion solution to the strong CP problem is automatically protected against the
quantum gravity corrections?  This question was pioneered in \cite{stanford}, where 
some wormhole solutions were found (generalizing earlier studies \cite{wormhole}), 
and suppression of amplitudes was estimated. 


  In the present article we shall discuss some related issues, but from somewhat different 
  perspective.  
  We shall first  establish an exact duality between the  $B_{\mu\nu}$-description and the 
  axion description with an arbitrary potential $V(a)$. 
The axion potential is entirely determined by the 
functional dependence of the Lagrangian on the three-form field strength, but not on its derivatives. 
As a result, in the presence of a single massless three-form, the minimum of the axion potential 
always coincides with the zero expectation value of the three-form field strength, irrespective of
the corrections.  That is, axion 
always screens the three-form electric field.  As said above,  this fact immediately explains why in QCD the axion solves the strong CP problem. Because the QCD three-form field strength is Tr$F\tilde F$, the axion 
minimum is always where $\langle$Tr$F\tilde F\rangle \, = \, 0$. 


 
  From here it follows that, in order to shift the axion minimum from the CP-conserving point, 
  any new physics  must un-Higgs the three-form gauge theory.  In $B_{\mu\nu}$-description 
  this is only possible if 
  there is an additional massless  three-form gauge field to which $B_{\mu\nu}$ could mix, or equivalently, an analogous massless
 correlator.   In the absence of such fields the axion solution is protected by 
the QCD gauge symmetry. 

It is known that gravitational or world-sheet instantons can contribute to the explicit breaking of 
PQ symmetry\cite{wenwitten}.  So one may wonder, how such contributions can be 
understood in the language of the three-form Higgs effect. In the light of the previous discussion,
one should be able to understand them as  un-Higgsing of the QCD three-form due to the 
presence of the additional  three-form fields. 

The obvious candidate for the additional massless three-form that can be provided by gravity
is the Chern-Simons spin connection three-form.
Due to  gravitational instantons, in the absence of additional anomalous symmetries, 
this field can provide the axion a mass and unscreen
the QCD three-form.  In the presence of the additional anomalous (with respect to gravity) symmetries,  such a contribution to the axion mass can be avoided.


In case when there are no additional massless three-forms available,  gravity can still generate a 
contribution to the axion mass,  but by gauge symmetry this corrections are 
automatically CP-conserving, because they cannot un-screen the  four-form electric field.

 We also generalize the discussion of dual three-form gauging to other  global symmetries, such as the baryon number. 
 For this we introduce a pseudo-scalar $b$ (bary-axion), which shifts under the baryon number symmetry.  Then,  by the reasons similar to  QCD,  baryon number becomes dual to a gauge symmetry gauged by an electroweak composite Chern-Simons three-form. 
  The difference from the axionic case 
 is that, because of $B+L$-symmetry,  bary-axion stays massless up to gravity corrections. 

\section{Three-Form Gauging of Axion Symmetries}

\subsection{Three-Form Higgs Effect} 

 We begin by considering the following  Lagrangian of a massless  pseudo-scalar field (an axion) $a$
 \begin{equation}
\label{freeaxion}
 L\, = \, f^2 (\partial_{\mu} \, a)^2,
\end{equation} 
where $f$ is an axion decay constant.  The above action is invariant under a continuous shift symmetry
\begin{equation}
\label{shift}
a\, \rightarrow \, a\,  + \, c,
\end{equation}
where $c$ is an arbitrary constant. 
Our goal is to create a mass gap in the above theory, by giving the axion a mass.  However,  in the same time we wish to satisfy the following two conditions

~~~~~

{\it 1)} Maintain the shift symmetry (\ref{shift}) (up to possible effect of the boundary terms); 

~~~~~

  {\it 2)} Do not increase the number of propagating degrees of freedom. 

~~~~

The first condition is usually satisfied when we deal with the gauge symmetries in the Higgs phase. 
Because gauge symmetries are the  redundancies of the description, they stay unbroken 
also in the Higgs phase,  despite the fact that
there are no massless excitations in the spectrum.  We shall use this fact as our guideline. However, the conventional  gauging of the shift symmetry (by a spin-1 vector  field) cannot satisfy the second requirement, because
it is accompanied  by additional propagating degrees of freedom.  Indeed, we could try to gauge the shift symmetry (\ref{shift}) by introducing a vector gauge field (a one-form) $A_{\mu}$. The Lagrangian 
then becomes
\begin{equation}
\label{higgs}
L \, = \, f^2(\partial_{\mu} a \, - \, A_{\mu})^2 \, - \, {1 \over 4} F_{\mu\nu}F^{\mu\nu}
\end{equation}
The global shift symmetry (\ref{shift}) is then promoted into a local gauge symmetry
\begin{equation}
\label{localshift}
a \, \rightarrow \, a \, + \, \alpha(x), ~~A_{\mu} \, \rightarrow \, A_{\mu} \, + \, \partial_{\mu}\, \alpha(x), 
\end{equation}
which for a particular case $\alpha \, =\, c$ reproduces (\ref{shift}). In the same time there are no massless excitations in the spectrum. Thus, the first condition is satisfied. However, the introduction 
of the vector gauge field adds the two additional degrees of freedom, and our second requirement is 
not fulfilled.  This happens because $a$ plays the role of the St\"uckelberg field for $A_{\mu}$, 
which enables to write a gauge invariant mass term for photon.  In reality what we want is rather having a 
St\"uckelberg field for $a$ itself, that would enable us to give it a mass in a shift-invariant way. 
The requirement of no new degrees of freedom restrict  such a "St\"uckelberg" field to being a 
gauge three-form $C_{\alpha\beta\gamma}$.
  
  In order to see how this works, it is  useful to use in parallel  
 a dual picture, in which the pseudoscalar $a$ is replaced by an antisymmetric Kalb-Ramond two-form 
 field $B_{\mu\nu}$.  The Lagrangian for a free $B_{\mu\nu}$-field is
 \begin{equation}
\label{RB}
P_{\alpha\beta\gamma} \,P^{\alpha\beta\gamma},
\end{equation}
where 
 \begin{equation}
\label{FB}
P_{\alpha\beta\gamma}\, = \, \partial_{[\alpha} B_{\beta\gamma]} \, = \, dB
\end{equation}
is the three-form field strength, and $d$ is the exterior derivative. 
This action is  invariant under a global shift symmetry  
\begin{equation}
\label{shiftB}
B_{\alpha\beta} \, \rightarrow  \, B_{\alpha\beta}  \, + \, \Omega_{\alpha\beta},
\end{equation}
where  $\Omega_{\alpha\beta}$ is an arbitrary  constant two-form.\footnote{ In addition there is also a 
gauge invariance under  $B_{\alpha\beta} \, \rightarrow  \, B_{\alpha\beta}  \, + \, \partial_{[\alpha}\xi_{\beta]}$, 
where $\xi_{\beta}$ is a one-form.}
We now wish to gauge (\ref{shiftB}) and create a mass gap. 
We shall achieve this by introducing a three-form gauge field $C_{\alpha\beta\gamma}$. 
However, before coupling it to $B_{\mu\nu}$, let us briefly recall some properties of a massless
three-form field. 
For a massless three-form field the lowest order parity-invariant Lagrangian
has the following form
\begin{equation}
\label{Caction}
 L\, =  \, F_{\mu\alpha\beta\gamma}F^{\mu\alpha\beta\gamma} \, +  \, C_{\alpha\beta\gamma}\, J^{\alpha\beta\gamma},
\end{equation}
where $F_{\mu\alpha\beta\gamma}\, = \, \partial_{[\mu}C_{\alpha\beta\gamma]}$ is the four-form
field strength, and  $J^{\alpha\beta\gamma}$ is a conserved external current
\begin{equation}
\label{conserved}
\partial_{\alpha}\, J^{\alpha\beta\gamma}\, = \, 0
\end{equation} 
The action (\ref{Caction}) is  then invariant under the gauge transformation
\begin{equation}
\label{gauge}
C_{\alpha\beta\gamma}  \rightarrow  C_{\alpha\beta\gamma} \, + \, d_{[\alpha}\Omega_{\beta\gamma]},
\end{equation}
where $\Omega$ is a two-form. 
Because of this gauge freedom  in four dimensions $C$ contains no propagating degrees of freedom.
Despite the absence of propagating degrees of freedom,  $C$ nevertheless can create a "Coulomb"-type long-range electric field in the vacuum $ F_{\mu\alpha\beta\gamma} =  F_0\epsilon_{\mu\alpha\beta\gamma}$. As it is obvious from the equation of motion, in the absence of sources, the four-form electric field can assume an 
{\it arbitrary} constant value. In the presence of the external  sources (which are two-branes or domain walls) the value of $F_0$ changes at the source.  
The actual presence of such sources for our
discussion is inessential.  

  We shall now couple $C$ to the two-form  $B_{\mu\nu}$ in order to gauge the shift symmetry
  (\ref{shiftB}). The simplest  Lagrangian has the following form
\begin{equation}
\label{CBaction}
 L\, = \,   12\, m^2 \, (P_{\alpha\beta\gamma}  \, - \, C_{\alpha\beta\gamma})^2 \, + \,  {1 \over 2} \, F_{\mu\alpha\beta\gamma}F^{\mu\alpha\beta\gamma}.  
\end{equation}
It is easy to see that this Lagrangian describes a Higgs phase of the  gauge theory invariant under (\ref{gauge}). As usual, in the Higgs phase the electric field is completely
screened in the vacuum. 
 For instance, the four-form electric
field created by a static probe brane source located at $z=0$ decays exponentially fast 
\begin{equation}
\label{screened}
F^{\mu\nu\alpha\beta} \,  \propto  \, {\rm sign}(z)\, {\rm e}^{-|z|m} \epsilon^{\mu\nu\alpha\beta}.
\end{equation}
Thus,  there are no massless excitations in the spectrum. However, because $C$ does not add
any new physical degrees of freedom the number  of propagating states is unchanged. 

 Because we have maintained the gauge symmetry of $B_{\mu\nu}$ and simultaneously give it a
 mass, we expect that the same property should persist in the dual axion description.  In order to see this explicitly, we perform the duality transformation  by treating $P_{\alpha\beta\gamma}$ as a fundamental 
 three-form, and enforcing  Bianchi identity
 \begin{equation}
\label{bianchi}
\epsilon^{\mu\alpha\beta\gamma} \partial_{[\mu}\, P_{\alpha\beta\gamma]}\, = \, 0, 
\end{equation}
 as a Lagrange constraint.  The new action takes the form 
\begin{equation}
\label{CBactiondual}
 L\, = \,  12 m^2 \, (P_{\alpha\beta\gamma}  \, - \, C_{\alpha\beta\gamma})^2 \, - \,  {1 \over 2} \, F_{\mu\alpha\beta\gamma}F^{\mu\alpha\beta\gamma} \,  +
 \,{\Lambda^2 \over 24} a\,\epsilon^{\mu\alpha\beta\gamma} \partial_{[\mu}\, P_{\alpha\beta\gamma]}   
\end{equation}
In this action, the axion $a$ appears as a Largange multiplier field.  Due to this, the axionic shift symmetry (\ref{shift}) on the equations of motion is guarantied.  Integrating out $P$ through its equation
of motion, we get the following dual action (up to total derivatives)
\begin{equation}
\label{aFaction}
L \, =  {\Lambda^4 \over 2m^2} \, (\partial_{\mu}a)^2 \,  -  \Lambda^2 \, \partial_{[\alpha} aC_{\beta\gamma\delta]}\epsilon^{\alpha\beta\gamma\delta}\, -  \, {1 \over 2} F_{\mu\alpha\beta\gamma}F^{\mu\alpha\beta\gamma}, 
\end{equation}
 Hence, the shift symmetry (\ref{shift}) is the automatic consequence of the gauge symmetry (\ref{shiftB}). 
 Because the gauge symmetry remains unbroken in the Higgs phase, so must be the global 
 shift symmetry in the dual description (up to boundary terms). 
 
 
   
   Integrating out $C$ through its equation of motion,  we get the following 
 effective equation of motion for the axion
\begin{equation}
\label{effa}
\Lambda^2\partial^2 a \,  + \, m^2 (a\, - \, \kappa) \, = \, 0.
\end{equation}
where $\kappa$ is an integration constant. 
\footnote{Because $\kappa$ is not a parameter in the Lagrangian but rather 
 an integration constant,  one could prescribe that under the shift symmetry (\ref{shift}) $\kappa$ changes as $k\, \rightarrow \, k \, + \, c$, in order to keep the shift symmetry manifest.  However, this prescription changes nothing physically, since (\ref{effa}) still described a single massive scalar degree of freedom.}

\subsection{Duality with Arbitrary Axion Potential}
 
We shall now  generalize the above duality to the case of an axion with an arbitrary potential and give a simple rule which relates this potential to the $F$-dependence of the Lagrangian.  
As discussed in the appendix, under the assumption of no additional massless fields the
axion potential is determined by the $F$-dependence of the action.  We therefore restrict our analysis here to a simplest gauge-invariant $B_{\mu\nu}$ action, that in the dual picture gives an arbitrary axion potential
(more general treatment is given in the appendix).  It is easy to see
that the following gauge-invariant theory
\begin{equation}
\label{bkaction}
 L_{B} \, =  {\Lambda^4 \over 24} \, {\bf K}\left({F \over \Lambda^2}\right) \, + \,  
 \,  12  m^2 \, (\partial_{[\alpha}B_{\beta\gamma]}  \, - \, C_{\alpha\beta\gamma})^2 \,,
 \end{equation}
where ${\bf K}\left({F \over \Lambda^2}\right)$ is an arbitrary function of its argument 
 $F\, =\, F_{\mu\alpha\beta\gamma}\, \epsilon^{\mu\alpha\beta\gamma}$, 
 is dual to 
\begin{equation}
\label{akaction}
L_a \, = \,   {\Lambda^4\over 2m^2} \, \partial_{\mu}a\partial^{\mu} a\, - \, \Lambda^2\, V(a ),
\end{equation}
where the axion potential is determined by the function ${\bf K}$ in the following way
\begin{equation}
\label{connection1}
{d V(a) \over da} \, = \, - \, F \, =\,-\, \Lambda^2 \, {\rm inv}{\bf K}'(\kappa \, - \, a).
\end{equation}
Here ${\rm inv}{\bf K}'(\kappa \, - \, a)$ is the inverse function of ${\bf K}'(\kappa \, - \, a)$, and
prime denotes a derivative with respect to the argument. 
$\kappa$ as before is an integration constant. 

 The duality can be established by starting from the action
\begin{equation}
\label{AFeffective}
 L\, =  {\Lambda^4 \over 24} \, {\bf K}\left({F \over \Lambda^2}\right) \, + \,  
 \,  12  m^2 \, (P_{\alpha\beta\gamma}  \, - \, C_{\alpha\beta\gamma})^2 \, + \,  {\Lambda^2\over 24}  \,a\,\epsilon^{\mu\alpha\beta\gamma} \partial_{[\mu}\, P_{\alpha\beta\gamma]},  
\end{equation}
and integrating out 
$P_{\alpha\beta\gamma}$ and $C_{\alpha\beta\gamma}$ through their equations of motion.
As a result of this integration one arrives to the following 
 effective equation for the axion
\begin{equation}
\label{axeffective}
   \partial_{\mu}\partial^{\mu} \, a\,+ \, 
  m^2 \, {\rm inv}{\bf K'}(\kappa \, - \, a)\, = \, 0
\end{equation}
Thus, the effective potential for the axion up to an arbitrary additive  constant 
\begin{equation}
\label{connection}
  V(\kappa\,  -\, a) \, = \, - \Lambda^2 \int \, {\rm inv}{\bf K}'(\kappa \, - \, a)\, da.
\end{equation}
For the simplest case ${\bf K}\left ({F \over \Lambda^2} \right)\, =\, {F\over 2\Lambda^2}$, we 
get (\ref{effa}).

Given $V(a)$ we can thus determine how the action  dependens on $F$ through (\ref{connection}). 
At the level of three-form gauge theory, all the information about the periodicity of the axion potential is encoded in the function ${\bf K}$.
For example, in QCD, where $V(a)$  is $2\pi$-periodic, in the dilute instanton gas approximation 
one may take  $V(a) \, =\, \cos(a-\kappa)$. This choice fixes the form of the ${\bf K}$-function
(for $\Lambda^2 =1$) as 
\begin{equation}
\label{kforcos}
{\bf K}(F) \, = \, F\, \arcsin(F) \, + \, \sqrt{1 \, - \, F^2}  
\end{equation}
An important feature guaranteed by (\ref{connection1}) is that irrespective of the form 
of ${\bf K}$ the four-form field strength automatically vanishes at the extrema of the axion 
potential.  This fact is just a dual version of the statement that $F$ is in the Higgs phase and 
therefore is screened in the vacuum.  As we shall see, this fact is the key point of the axion 
solution of the strong CP problem in QCD, because in QCD the expectation value of
$F$ is equal to the one of Tr$F\tilde F$. 

\subsection{Multiple Three-forms}

The above duality can be generalized to multiple three-forms. For instance, 
in the case of two three-forms $C_1$ and $C_2$ the theory
\begin{equation}
\label{bkaction12}
 L_{B} \, =  {\Lambda_1^4 \over 24} \, {\bf K_1}\left({F_1 \over \Lambda_1^2}\right) \, + \,  
 {\Lambda_2^4 \over 24} \, {\bf K_2}\left({F_2 \over \Lambda_2^2}\right) \, +
 \,  12  m^2 \, ( dB  \, - \, C_1 \, -\, C_2)^2 \,,
 \end{equation}
 is dual to 
\begin{equation}
\label{akaction12}
L_a \, = \,   {\Lambda^4\over m^2} \, \partial_{\mu}a\partial^{\mu} a\, - \, \Lambda^2\, (V_1(a\, - \,\kappa_1) \, + \, V_2(a \, - \, \kappa_2)),
\end{equation}
where ($j = 1,2$)
\begin{equation}
\label{connection12}
{d V_j(a\, - \kappa_j) \over da} \, = \, - \, F_j \, =\,-\, \Lambda_j^2 \, {\rm inv}{\bf K_j}'\left ({\Lambda^2\over \Lambda_j}(\kappa_j \, - \, a)\right).
\end{equation}
The important difference from the single three-form case is that the resulting theory (\ref{akaction12})
has a continuum of physically-inequivalent vacua that are scanned by the difference of the two 
integration constants $\kappa \, = \, \kappa_1 \, - \, \kappa_2$. These vacua obey the superselection rule
and are very similar to $\theta$-vacua of QCD.  Their existence is the consequence of the fact that 
a single axion cannot Higgs more than one three-form field.  Therefore, the orthogonal combination
remains in the Coulomb phase, and the superselection vacua are simply the vacua with different
values of the unscreened four-form electric field  $F_1 \, - \, F_2$.  We shall encounter this effect later
in the context of the strong CP problem.  

 The periodicities of the two functions $V_1$ and $V_2$ can in principle be different. 
 In such a case, within each superselection sector (fixed $\kappa$-s), one gets a family of vacua 
 with different energies.  Such models may find  application in 
 theories with statistical vacuum selection.
 For instance, we can have $V_1\, = \, \Lambda_1^2 a$
 and $V_2\, = \, \Lambda_2^2 {\rm cos}(a)$ with $\Lambda_1^2 << \Lambda_2^2$, in which case the axion potential will look as "washboard" potential obtained in (\cite{giaalex}).  
 
  If the difference of the periodicities is  irrational, the above theory gives a simple realization of 
  the irrational axion idea\cite{irax}.

\section{The Strong CP Problem and its Solution in Three-Form Language}


We now wish to reformulate the strong CP problem in the language of a three-form gauge theory. 

We shall do this first in a simple model with a fundamental three-form field $C_{\alpha\beta\gamma}$. In this theory a full equivalent of the strong CP problem does exist, and this problem is solved by axion. 
Then we discuss the real QCD with axion, which also can be reformulated  in the language 
of a three-form gauge theory.  In both cases the strong CP problem and its axion solution has an universal 
formulation in terms of the Coulomb and the Higgs phases of a three-form gauge theory. 
Any theory in which a three-form field is in the Coulomb phase 'suffers' from a generalized strong CP problem.  The axion  solution to this problem is then nothing but  the Higgsing of the three-form
gauge field. 

Consider a theory of a massless three-form field, with the simplest Lagrangian (\ref{Caction}). 
This theory is in the Coulomb phase, and this fact is the source of the generalized strong CP problem. 
The equations of motion are solved by an arbitrary constant electric field
\begin{equation}
\label{celectric}
F^{\mu\nu\alpha\beta} \, = \,F_0\,  \epsilon^{\mu\nu\alpha\beta},
\end{equation}
where $F_0$ is arbitrary, and plays the same role as the $\theta$-parameter in QCD.  
In particular, the constant electric field (\ref{celectric}) is CP-odd.  Also, there is a super-selection rule among the
$F_0$-vacua, very similar to the one of $\theta$-vacua in QCD.  Note that, we could have introduced
the $\theta$-parameter directly in the Lagrangian through the following term
\begin{equation}
\label{thetaf}
\theta \, F,
\end{equation}
but this would just redefine the value of the constant electric field. 
Hence, the $\theta$-parameter in a three-form gauge theory is equivalent to a constant 
four-form electric field in the vacuum. 
Thus, the strong CP problem reformulated in the language of a three-form gauge theory, reduces to the following question.
How can the four-form Coulomb electric field be made naturally small? In such a formulation the answer is immediately obvious. There is a well known way to get rid of any constant electric field in the vacuum (i.e. to screen it), 
which is putting the gauge theory in the Higgs phase. This is achieved by introduction of a $B_{\mu\nu}$-axion, transforming under 
the gauge shift symmetry  (\ref{shiftB}). The simplest  Lagrangian accomplishing this is (\ref{CBaction}),
and this solves the strong CP problem. 
As was shown above, the electric field becomes screened (\ref{screened}). 
As we see, in the $B_{\mu\nu}$ language the axion solution to the strong CP problem has a very simple physical interpretation.  In the dual $a$-axion language, the reason for screening is translated as the fact that the minimum of the axion potential $V(a)$ is always at the point where $F=0$. This is an automatic
consequence of the universal relation (\ref{connection1}),  which  implies
\begin{equation}
\label{vofa}
\langle F  \rangle \, = \,-\,  \Lambda^2 {d V(a) \over da},
\end{equation} 
 Now we are ready to ask the question, how could gravity (or any other physics) 
 undo the above solution to the strong CP problem? 
 In the language of axion $a$ this would correspond to shifting the axion VEV away from the 
$F=0$ point, that is, undoing the relation (\ref{vofa}).  This would mean that at the minimum of the axion 
potential the four-form electric field is no longer 
screened and the  three-form gauge theory is back to the Coulomb phase. 

 It is simplest to answer this question  in $B_{\mu\nu}$-language. 
There is an unique gauge-invariant way for  un-screening a gauge field,  which is to get rid of the Goldstone-St\"uckelberg  field
by making it massive.  Thus, $C_{\alpha\beta\gamma}$ can be taken out of the Higgs phase 
if we could make $B_{\mu\nu}$ independently-massive. 
 But by gauge symmetry, this can only be achieved by introducing an additional massless three-form field, to which $B_{\mu\nu}$ could mix.  We shall call this new field  $G_{\alpha\beta\gamma}$.

 At the end of the previous section we have already seen how in the presence of an additional
 three-form the electric field cannot be completely screened by the axion. 
 It is obviously necessary that the axion also shifts
 \begin{equation}
\label{shiftB2}
B_{\alpha\beta} \, \rightarrow  \, B_{\alpha\beta}  \, + \, \Upsilon_{\alpha\beta},
\end{equation}
under the second gauge symmetry 
\begin{equation}
\label{gauge2}
G_{\alpha\beta\gamma}  \rightarrow  G_{\alpha\beta\gamma} \, + \, d_{[\alpha}\Upsilon_{\beta\gamma]},
\end{equation}
where $\Upsilon$ is a two-form. 
 
  Then the $B_{\mu\nu}$-axion 
 will be eaten up 
 by one combination of $C$ and $G$, whereas  the orthogonal combination will remain in the Coulomb phase.  The 
simplest  Lagrangian accomplishing this goal is 
\begin{equation}
\label{CGaction}
 L\, = \,  12 \, m^2 \, (dB \, - \, C\, -\, G )^2 \, + \,  {1 \over 48} \, F^2 \, +   {1 \over 48} \, (F^{G})^2,
\end{equation}
where $F^G \, = \,  F^{G}_{\mu\alpha\beta\gamma}\epsilon^{\mu\alpha\beta\gamma},$ is the field strength of of $G$.  As a result, one combination 
of the electric fields $F \, -\, F^G$ is no longer screened
and can assume an arbitrary constant values. Thus, the strong CP problem got reintroduced. 
 Note that unscreening of the $F$-field is equivalent of violating the relation (\ref{vofa}).
Indeed,  dualizing (\ref{CGaction}), we can see how the problem looks in the $a$-language. 
 In this dual language the manifestation of the problem is that  the relation (\ref{vofa}) no longer holds
 and in accordance with (\ref{akaction12}) and (\ref{connection12}) the axion  potential is instead given by 
 \begin{equation}
\label{vofanew}
F\, + \, F^G \, = \,- \, \Lambda^2\,  {dV(a) \over da} 
\end{equation}
Hence, the minimum of the axion potential no longer corresponds to the point where  the 
electric field vanishes.  Note that  the possibility  of unscreening the electric field is completely independent
of the concrete form of the action, but is only determined by the degrees of freedom available in the theory. 
Only if  there is a second three-form field, and a single axion, the electric field will be 
partially unscreened.  In more general terms, the number of the three-form fields 
must  exceed the number of axions. 
 
 Note that the above finding is in full agreement with the appendix result, according to which the
 appearance of the additional massless poles in the effective action of $C_{\alpha\beta\gamma}$ is the only way in which 
  the relation (\ref{vofa}) may be violated. 
 Indeed, integrating out the $G_{\alpha\beta\gamma}$ in (\ref{CGaction}) we get 
 \begin{equation}
\label{intG}
L\, =\, 12\,  A_{\alpha\beta\gamma} {m^2\partial^2 \over \partial^2 \, +\, m^2} \Pi^{[\alpha}_{\mu}  A^{\mu\beta\gamma]} \, + \, {1 \over 48} F^2 ,
\end{equation}
where $A_{\alpha\beta\gamma}\, = \,  P_{\alpha\beta\gamma} - C_{\alpha\beta\gamma}$ and 
 $\Pi_{\mu\nu} \, = \, \eta_{\mu\nu} \, - \, \partial_{\mu}\partial_{\nu}$. This expression is equivalent to (\ref{lexample}) with  ${\bf O}(\partial^2) \, = \,  {m^2\partial^2 \over \partial^2 \, +\, m^2} $. 

 Thus, in order  for any new physics to undo the axion solution of the above strong CP problem,
it must provide an independently-massless three-form field.

\section{QCD Axion in Three-Form Language}

  Let us now  discuss the axion solution of the strong CP problem in QCD, and show that 
it can be  reformulated in the above-presented thee-form language.  

  
 In order to translate the QCD strong CP problem in the three-form language,  consider a $\theta$-term in  $SU(N)$ gauge theory with a strong coupling scale $\Lambda$ (which we shall set equal to one) and no
light fermion flavors 
\begin{eqnarray}
\label{apsi}
L \, = \, \theta\, {g^2\over 32\pi^2}
 F^a\tilde{F^a}, 
 \end{eqnarray}
where $g$ is the gauge coupling,  $\tilde{F^a}$ is a dual field strength and $a$ is an  $SU(N)$-adjoint index.
 We first note that  this term can be rewritten as a four-form fields strength $F$ of a composite three-form
 \begin{eqnarray}
\label{apsi}
L \, = \, \theta\, {g^2\over 32\pi^2} F^a\tilde{F^a} \, = \, \theta \, F\, = \, \theta\, F_{\alpha\beta\gamma\delta}
\epsilon^{\alpha\beta\gamma\delta}, 
 \end{eqnarray}
where
\begin{equation}
\label{qcdform}
C_{\alpha\beta\gamma} \, = \,  {g^2\over 8\pi^2}\, {\rm Tr} \left (A_{[\alpha}A_{\beta}A_{\gamma]}\,  - {3 \over 2}
A_{[\alpha}\partial_{\beta}A_{\gamma]}\right ),
\end{equation}
is the Chern-Simons three-form and where $A_{\alpha} \, = \, A^a_{\alpha}T^a$ is the gauge field matrix, and $T^a$ are the generators of the group.  Under the gauge transformation, $C$ shifts as (\ref{gauge}) with
\begin{equation}
\label{omega}
\Omega_{\alpha\beta}\, = \, A^a_{[\alpha}\partial_{\beta]}\omega^a
\end{equation} 
where $\omega^a$ are the $SU(N)$ gauge transformation parameters. The four-form field strength
\begin{equation}
\label{F}
F_{\mu\alpha\beta\gamma}\, = \, \partial_{[\mu}C_{\alpha\beta\gamma]} 
\end{equation}
is of course invariant under (\ref{gauge}) and (\ref{omega}). 
Note that the $SU(N)$  Chern-Simons current $K_{\mu}$ can be written as 
\begin{equation}
\label{cs}
K_{\mu} \, = \, \epsilon_{\mu\alpha\beta\gamma} C^{\alpha\beta\gamma} 
\end{equation} 
It is known \cite{qcdform} that at low energies, the three-form $C$ becomes a massless {\it field}, 
and creates a long-range Coulomb-type constant force. 
The easiest way to see that $C$ mediates a long-range interactions is through the 
Kogut-Susskind pole \cite{KS}.   The zero momentum limit of the  following correlator  
\begin{equation}
\label{corr}
 {\rm lim} _ {q \rightarrow 0} \, q^{\mu}q^{\nu} \, \int d^4x  {\rm e}^{iqx} \langle 0| T K_{\mu}(x)K_{\nu}(0)|0\rangle
\end{equation}
is non-zero,  as it is related to topological susceptibility of the vacuum, which is 
a non-zero number in pure gluodynamics.  Hence, the correlator of the two  Chern-Simons
currents has a pole at zero momentum, and the same is true for the correlator of 
two three-forms.  Thus,  the three-form field develops a Coulomb propagator and mediates a  
long-range force.  Because the probe sources for the three-form are two dimensional surfaces 
(domain walls or the two-branes), the force in question is constant.  

 In the other words,  at low energies, the QCD Lagrangian contains a massless three-form field, and can be written as 
 \begin{equation}
\label{qcdtheta}
L \, = \, \theta F \, + \, {\bf K}(F)\, + \, ...,
\end{equation}
The exact form of the function ${\bf K}$ in QCD is unknown, but the strong CP problem is solved 
by axion regardless of this form. 
It is obvious now that the $\theta$-problem in  QCD is isomorphic to a problem of a constant 
four-form electric field, and that QCD $\theta$-vacua are simply  vacua with different values of this electric field. 

 Axion solution of this problem is nothing but Higgsing this composite three-form given in 
 (\ref{qcdform}).
That is, the axion solves this problem by giving a gauge-invariant mass to the three-form  field and screening it in the vacuum.  Indeed, the axion solution is based on the idea of promoting 
$\theta$ into a dynamical field $a$, which immediately gives us the generalized version 
of the Lagrangian  (\ref{aFaction})
\begin{equation}
\label{aFactionqcd}
L \, =  {f^2 \over 2} \, (\partial_{\mu}a)^2 \,  - \,  a \,  F \, -  \,{1 \over 24} {\bf K}(F) 
\end{equation}
which is dual to (\ref{bkaction}).  After integrating over $F$ this gives an effective axion equation
(\ref{axeffective}) in which the function ${\bf K}$ determines  the axion potential through
(\ref{connection1}). 
Because of the relation (\ref{apsi}) the  relation (\ref{vofa}) guarantees that 
 $\langle F^a\tilde{F^a} \rangle \, = \, 0$ in axionic vacuum, and the strong CP problem is solved.  
 Note that, the relation (\ref{connection}) reproduces the well known relation
 in QCD 
 \begin{equation}
\label{thetarel}
{d E(\theta) \over d \theta} \, \propto \, \langle F\tilde F \rangle
\end{equation}
 As it is clear from our previous discussion, in solving the strong CP problem we can use exclusively 
 $B_{\mu\nu}$-language.  This language has an advantage that we introduce PQ symmetry as 
 intrinsically-gauge symmetry (\ref{gauge}) with the gauge parameter set by (\ref{omega}).  In order to accomplish this, all we have to do is to 
 add to the QCD Lagrangian the following gauge invariant mass term analogous to the one in (\ref{CBaction}) and (\ref{bkaction})
 \begin{equation}
\label{masstrem}
(\partial_{[\alpha}B_{\beta\gamma]}  \, - \, C_{\alpha\beta\gamma})^2,
\end{equation}
 where under $C_{\alpha\beta\gamma}$ we have to understand (\ref{qcdform}).  Obviously, the same 
mass term can be obtained by dualizing the action (\ref{aFactionqcd}).  The addition of such a mass term screens the QCD four-form field (\ref{F}) \cite{screen}.
We see that this screening is the essence of the solution of the strong CP problem. 
By duality the three-form Higgs effect is equivalent to the axion minimum being at zero $\theta$.
 
  The three-form language makes the connection between the axion and the  massless quark solutions
 of the strong CP problem especially clear.   Indeed, in the case of the massless quark solution  there is an  axial $U(1)_A$-symmetry, which is spontaneously broken by the quark condensate. In the 
 absence of  the anomaly, there would be a massless Goldstone boson, the $\eta'$-meson. 
This boson gets a mass from non-perturbative QCD effects due to $U(1)_A$-anomaly, according to
't Hooft's  solution of the $U(1)$-problem\cite{tsolution}. In the above three-form language these non-perturbative effects amount to Higgsing the QCD three-form field. This field eats up the $\eta'$ meson
and gives it a mass. As a result of this effect the four-form electric field gets screened and the strong
CP problem is solved.  That is, in the case of massless quarks $\eta'$-meson plays the same 
role as the axion in Higgsing the three-form field. 
 Note that in the case of massless  quarks, substituting $a$ with $\eta'$, the relation (\ref{connection1}) 
 implies the Witten-Veneziano relation for the $\eta'$ mass\cite{wittenveneziano}. 

\section{Gravity} 

 Non-perturbative quantum phenomena, such as  gravitational or string world-sheet instantons\cite{wenwitten} could undo the axion solution to the strong CP problem.  We now wish to understand, how such dangerous corrections could manifest themselves in the 
three-form  gauge theory language. 
    
  First of all, we have seen that the strong CP problem is equivalent to the problem of a constant 
four-form electric field in QCD vacuum.  The axion solution to this problem is nothing but the 
three-form Higgs
effect.  The  $B_{\mu\nu}$-axion is eaten-up by a three-form, and the electric field    
is screened.  This is why, in the absence of the additional massless three-forms, the minimum of the axion
potential is {\it always} at the point where $F$-vanishes, as it is enforces by the the relation
(\ref{connection1}) or (\ref{vofa}). 

   Thus,  gravity can only jeopardize the  axion solution of the strong CP problem, if it can
  {\it unscreen} the electric $F$-field. That is, the new physics must take the three-form gauge theory
  out of the Higgs phase back to the Coulomb phase.
 In this respect, it is important to distinguish the harmless corrections that correct axion potential without 
 unscreening the QCD  four-form, from the dangerous ones, that can unscreen it.  
 
 The first type of corrections are very easy to imagine.  In fact any physics that corrects the 
function ${\bf K}(F)$ without introducing some additional massless three-forms, will be harmless. 
As long as there is only one three-form $C$ and a single two-form $B$, the gauge symmetry will 
remain in the Higgs phase irrespective of the form of ${\bf K}(F)$-function.

 Obviously,  the same conclusion holds true in the dual picture. 
  Any physics that corrects the form of the function ${\bf K}(F)$, will also change the axion potential according to (\ref{vofa}). So that the minimum of the corrected axion potential will 
continue to be at $F\, = \, 0$.   In the other words, 
the ${\bf K}$-function in equation (\ref{vofa}) can be viewed as the original QCD function plus arbitrary 
quantum gravity corrections 
\begin{equation}
\label{kcorrections}
{\bf K} \, =\, {\bf K}_{QCD} \, +\, {\bf K}_{quantum~gravity}\, +\, ...,
\end{equation}
but the electric $F$-field remains screened for an  arbitrary form of ${\bf K}$. 

 Now we wish to address the issue of the dangerous corrections, the ones that could take the three-form 
 theory out of the Higgs phase.  
   The three-form gauge theory is by no means unique
in this respect. The similar question could be asked about any 
 gauge theory that is in the Higgs phase, and there is an universal answer.  The only way to move a
 gauge theory from the Higgs to the Coulomb phase in a gauge-invariant way is to give an {\it independent} mass to the Goldstone-St\"ckelberg field by mixing it with an additional massless gauge field.   For example, if we wish to un-Higgs  the $U(1)$ vector gauge field $A_{\mu}$ in the toy example given by Lagrangian (\ref{higgs}), we have to introduce an additional  vector gauge field $G_{\mu}$  and mix it with $a$
 \begin{equation}
\label{Ghiggs}
L \, = \, f^2(\partial_{\mu} a \, - \, A_{\mu} \, -\, G_{\mu})^2 \, - \, {1 \over 4} F_{\mu\nu}F^{\mu\nu}
-\, {1 \over 4} F_{\mu\nu}^GF^{G\mu\nu}.
\end{equation}
Note the complete analogy with (\ref{CGaction}).
Equivalently,   we could have introduced a derivative interaction leading to the massless poles
in the Greens function of $A_{\mu}$
  \begin{equation}
\label{Gintegration}
L \, = \, f^2(\partial_{\mu} a \, - \, A_{\mu}) {\partial^2 \over \partial^2 \, + \, m^2} \,(\partial_{\mu} a \, - \, A_{\mu}) \,  - \, {1 \over 4} F_{\mu\nu}F^{\mu\nu}
\end{equation}
However, this is {\it not} an alternative way, but  the exact equivalent of (\ref{Ghiggs}), since 
(\ref{Gintegration}) can be obtained from (\ref{Ghiggs}) by integrating out $G_{\mu}$ field through 
its equations of motion.  Again, (\ref{Gintegration})  is a complete analog of (\ref{intG}).

 Thus, in $B_{\mu\nu}$-language the only way in which the gravity corrections  could undo the relation  (\ref{vofa}),  is  if 
 gravity could provide an additional massless three-form field $G_{\alpha\beta\gamma}$, which would contribute into the  $B_{\mu\nu}$  mass. Such an additional contribution to the axion mass
can only come from the gauge-invariant mass term of the form (\ref{CGaction}), or equivalently through the massless pole in the three-form propagator that may be provided by action(\ref{intG}). 
 
  After dualization, in $a$-language,  such dangerous corrections would translate 
into the mixing to an additional  three-form
\begin{equation}
\label{affg}
a\, (F\, + \, F^G).
\end{equation}
 
 The obvious candidate for an additional three-form in pure gravity theory is the Chern-Simons three-form
  (antisymmetrization  over $\alpha\beta\gamma$ is assumed)
 \begin{equation}
\label{chern}
G_{\alpha\beta\gamma}\, = \, {1\over 12} \Gamma^i_{j\alpha} \partial_{\beta} \Gamma_{i\gamma}^j
\, + \, {1 \over 18} \Gamma^i_{j\alpha} \Gamma_{k\beta}^j \Gamma_{i\gamma}^k,
\end{equation}
where $\Gamma$-s  are connections. 
That (\ref{chern}) is the right three-form follows from the fact that 
the topological invariant  $R\tilde{R}$ can be rewritten as the field strength of   
 $G$
 \begin{equation}
\label{rrtilde}
R\tilde R \, = \, \epsilon^{\alpha\beta\mu\nu}R^i_{j\alpha\beta}R^j_{i\mu\nu} \, =\, {1\over 3}
\, \epsilon^{\alpha\beta\mu\nu}\partial_{[\alpha}G_{\beta\mu\nu]} 
\end{equation}
 Under general coordinate transformations $G$ acquires an additional shift 
 by an exterior derivative of a two-form $\partial_{[\alpha}\Upsilon_{\beta\gamma]}$. 
 (This follows from the  Poincare lemma and the fact that (\ref{rrtilde}) is invariant). 
 Thus, if we demand that $B_{\mu\nu}$ shifts as (\ref{shiftB2}) with the same $\Upsilon$, the
 following field strength 
\begin{equation}
\label{dbinvariant}
 H\, = \, \partial_{[\alpha}B_{\beta\gamma]}  \, - \, C_{\alpha\beta\gamma}\, -\, G_{\alpha\beta\gamma} 
\end{equation}
will   be invariant under both gauge symmetries. 
This expression is identical to the gauge invariant field strength of the antisymmetric two-form
$B$-field  in ten dimensional string theory.  The above form there is dictated by 
Green-Schwarz
anomaly cancellation\cite{gsanomaly}, whereas in the present case it is generated by anomaly. 
Indeed through gravitational anomaly \cite{gravianomaly}, the axion acquires a
 coupling 
\begin{equation}
\label{arrtilde}
a\, R\tilde R
\end{equation}
and the expression (\ref{dbinvariant}) can be obtained by dualization after taking into the account (\ref{arrtilde}) coupling.

 In order for the coupling to the gravitational three-form $G$ to undo the axion solution of the strong
 CP problem, $G$ must provide a contribution to the axion mass.  This is possible, if 
 $G$ actually behaves as a massless three-form field, that is if  the propagator 
 of $G$ develops  a massless pole, analogous to its QCD counterpart (\ref{corr}).  This pole, can serve as an indication that underlying gravitational physics can lead to strong CP violation.  
 One way to avoid such a contribution is to have additional global symmetries that carry gravitational 
 anomaly.  This can be, for instance,  the Standard Model baryon plus lepton number symmetry, or some additional 
 symmetry from the sector that only couples through gravity. For instance, a sterile massless fermion
 will exhibit such an anomalous chiral  symmetry.  In the presence of such a symmetry, 
 gravitational three-form will not be able to contribute into the axion mass. This follows from 
 the following consideration.  It is well known\cite{gravianomaly2, gravianomaly1, gravianomaly}, that  in the presence of an anomalous global symmetry, the gravitational
 anomaly leads to a non-conservation of the corresponding current 
 \begin{equation}
\label{ganomaly}
\partial_{\mu} J^{\mu} \, \propto \, R \tilde R 
\end{equation} 
 As a result, the gravitational analog of the $\theta$-term can be rotated away by chiral transformation, much in the same way as  the $\theta$-term in nonabelian gauge theories in the presence of massless
 fermions.  Thus, in the presence of chiral symmetry the expectation value of the axion 
 cannot be observable through the coupling (\ref{arrtilde}), and thus, the latter cannot contribute to the 
 axion potential. 

  If there are no  additional chiral symmetries, one could still try to use the standard model 
  $B+L$ symmetry to rotate away the gravitational $\theta$-term. However, because $B+L$
  symmetry carries the electroweak anomaly, one will be left with the electroweak $\theta$-term. 
  So because one cannot simultaneously rotate away both $\theta$-terms, in the absence of an additional 
  chiral symmetry, the axion will get an extra contribution to its potential, which will unscreen the QCD three-form
  $C$. However, the effect will be suppressed by the strength of both gravitational and electroweak instantons.  So the net  effect may be sufficiently small  in order not to pose any danger for the axion solution. 
  
 However, if one would like to maintain an exact  CP invariance in QCD sector, an obvious way would be to introduce an additional two form field  $B'_{\mu\nu}$ in order to match the number of "dangerous" 
 three-forms.  $B'_{\mu\nu}$ should couple to different combination of the three-forms.  The QCD three-form then will remain in the Higgs phase, and the strong CP problem will be solved.


\section{Baryon number}

In this section we shall discuss a dual gauging of other global symmetries
such as baryon and lepton numbers.
 
 First we shall discuss a simplified  example, which 
we shall later extend to a fully realistic Standard Model.  The simplified model is a theory of a single 
Weyl (or a Majorana)   fermion $\psi$, which we shall call 'baryon", and which transforms under a global 
phase symmetry 
\begin{equation}
\label{bnumber}
\psi \, \rightarrow \, {\rm e}^{iq\alpha} \psi,
\end{equation} 
which we shall refer to as the  "baryon number" (or $B$-number, for short). Correspondingly, $q$ is the baryonic charge of $\psi$. 
We now wish to dualize  the baryon number symmetry into a symmetry gauged by a three-form field, but for this we have to assume that 
(\ref{bnumber}) is non-linearly realized. This can be achieved by introducing a Goldstone boson,
$b$, which shifts under (\ref{bnumber}) as 
\begin{equation}
\label{bshift}
b\, \rightarrow \, b \, + \, \alpha.
\end{equation}
We shall call $b$ a {\it bary-axion}.  
The Lagrangian with a three-form gauged baryon number symmetry can be written as 
\begin{equation}
\label{baction}
L \, = \, i \bar{\psi}\partial_{\mu}\gamma^{\mu} \psi\, + \, \mu{\rm e}^{ib}(\psi\, c\, \psi)^{{1\over 2q}}\, + \,   
\, {f^2 \over 2} \, (\partial_{\mu}b)^2 \,  + \, {\Lambda^2 \over 144} b \,  F  \, +  \,{\Lambda^2\over 24} {\bf K}(F), 
\end{equation}
where $c$ is the charge conjugation matrix.
 $\mu$ is the constant that sets the strength of the $B$-violating interaction. In cases, in which 
the baryon number is a linearly realized symmetry above a certain scale, 
$\mu$ can be identified with the VEV of a scalar that breaks $U(1)_B$ spontaneously. However,
existence of such a scalar is not essential for our purposes, and $U(1)_B$ may very well be a
non-linearly realized symmetry  at the field theory level.  
 By chiral rescaling of the baryon fields 
\begin{equation}
\label{brescale}
\psi \rightarrow \psi' \, = \, {\rm e}^{ib q}\psi
\end{equation}
we can eliminate all the non-derivative interactions of $b$  from the Lagrangian, which now takes the form  
\begin{equation}
\label{bactionrescale}
L \, = \, i \bar{\psi'}\partial_{\mu}\gamma^{\mu} \psi'\, + \, \mu (\psi'\, c\, \psi')^{{1\over 2q}}\, + \,   
\partial_{\mu}b\, J^{\mu}\, + 
\, {f^2 \over 2} \, (\partial_{\mu}b)^2 \,  + \, {\Lambda^2 \over 144} b \,  F  \, +  \,{\Lambda^2 \over 24} {\bf K}(F), 
\end{equation}
where $J^{\mu} \, = \, \bar{\psi}\gamma^{\mu}\psi$ is the baryonic current.   
Dualizing (\ref{bactionrescale}) we get the following action for the antisymmetric two-form $B_{\mu\nu}$
\begin{eqnarray}
\label{bactiondual}
L \, = \, i \bar{\psi'}\partial_{\mu}\gamma^{\mu} \psi'\, + \, \mu (\psi'\, c\, \psi')^{{1\over 2q}}\,   
- \, {J_{\mu}J^{\mu} \over 2f^2} \, + 
\,  {\Lambda^2 \over 6f^2} \, (\partial_{[\alpha}B_{\beta\gamma]}  \, - \, C_{\alpha\beta\gamma}) 
\epsilon^{\alpha\beta\gamma\mu}J_{\mu} \, \\ \nonumber
+ \,   {\Lambda^4 \over 12f^2} (\partial_{[\alpha}B_{\beta\gamma]}  \, - \, C_{\alpha\beta\gamma})^2\,
 +  \,{\Lambda^2 \over 24} {\bf K}(F), 
\end{eqnarray}

If an additional $B$-violating axion potential $V(b)_1$ is included in (\ref{baction}) the system will respond by unscreening the $F$-field. This effect on the $B_{\mu\nu}$-side can be accounted
by integrating in an additional three-form field $C_1$ with the appropriate kinetic function.

  We now wish to generalize the above construction to a fully realistic Standard Model. 
 The good news is that 
 we no longer  need an elementary  three-form, since  a non-linearly realized 
 baryon number symmetry in the Standard Model is dual to the symmetry  gauged by the composite Chern-Simons three-form of the electroweak gauge sector,  much in the same way as the PQ symmetry is gauged 
 by a composite QCD three-form (\ref{qcdform}).  
 
In order to promote the baryon number symmetry of the Standard Model into a non-linearly realized shift symmetry, 
 we add the following interactions to the usual Standard Model
Lagrangian 
\begin{equation}
\label{baction}
L \, = \, \mu{\rm e}^{ib}(QQQL) \, + \,   
\, {f^2 \over 2} \, (\partial_{\mu}b)^2, 
\end{equation}
where $Q$ and $L$ stand for quarks and leptons respectively.  For definiteness we have used the lowest possible baryon number carrying operator $QQQL$, but any other operator would do the job
equally well. The above action is invariant under the baryon number symmetry 
\begin{equation}
\label{bofq}
Q\rightarrow {\rm e}^{-i\alpha/3}Q, ~~ b \, \rightarrow \, b\, + \, \alpha
\end{equation}
For definiteness we shall assume that the lepton number is explicitly broken by neutrino masses, and 
will not consider it. However, this analysis can be trivially extended to any anomalous combination 
of $B$- and $L$-symmetries.

  Now, by rescaling quarks by a chiral transformation
\begin{equation}
\label{qrescale}
Q\rightarrow Q'\, =\, {\rm e}^{i b/3}Q,
\end{equation}
we eliminate all the non-derivative bary-axion interactions. The resulting Lagrangian is 
\begin{eqnarray}
\label{baction}
L \, = \, L_{SM}(Q') \, + \,  \mu \, (Q'Q'Q'L) \, + \,   \partial_{\mu}b J^{\mu}\, +  \, 
b{g^2\over 32\pi^2}
 F^a\tilde{F^a}\, + \, 
\, {f^2 \over 2} \, (\partial_{\mu}b)^2, \, \\ \nonumber
\end{eqnarray}
where $L_{SM}(Q')$ is the Standard Model Lagrangian in which $Q$-s are replaces by $Q'$-s, and
$J^{\mu}$ is the usual baryon number current. 
$F^a$ is an $SU(2)$ gauge symmetry field strength, and coupling of the bary-axion to it appears as a result of a baryon number anomaly.  As in the case of QCD, this coupling can also be rewritten 
as the coupling to a composite three form field,
 \begin{eqnarray}
\label{bweakf}
L \, = \, b\, {g^2\over 32\pi^2} F^a\tilde{F^a} \, = \, b \, F^{SU(2)}_{\alpha\beta\gamma\delta}
\epsilon^{\alpha\beta\gamma\delta}  
 \end{eqnarray}
where $F^{SU(2)}$ is the field strength of the following three-form 
\begin{equation}
\label{weakform}
C^{SU(2)}_{\alpha\beta\gamma} \, = \,  {g^2\over 32\pi^2}\, {\rm Tr} \left (A_{\alpha}A_{\beta}A_{\gamma}\,  - {3 \over 2}
A_{[\alpha}\partial_{\beta}A_{\gamma]}\right ),
\end{equation}
where $A_{\alpha} \, = \, A^a_{\alpha}T^a$ is the $SU(2)$ gauge field matrix, and $T^a$ are the generators of the group. This is in the full analogy with (\ref{apsi}) and (\ref{qcdform}), but the 
crucial difference from QCD in case of electroweak $SU(2)$ is that, there is no long-range field associated with $C^{SU(2)}$.  The consequence of this fact is that 
in  the absence of an additional massless three-form $C$, the bary-axion would remain exactly massless. 

 The masslessness of the bary-axion also follows from the fact that the energy is independent on the value of the electroweak  $\theta$-term. This was shown by explicit calculation  
 in ref\cite{anselm}.  A simple argument showing the same is that in the Standard Model  there  is an anomalous chiral symmetry $B+L$, which can rotate away the electroweak $\theta$-term and thus, the latter is unobservable.  However, as discussed above, this would be true only in the absence of
 gravity. In case of gravity, because of gravitational anomaly, one cannot get rid of both 
 $\theta$-terms by the same rotation.  So in the absence of additional chiral symmetries, 
 bary-axion would get a small mass.

 The fact that the bary-axion gets no contribution to its mass from the electroweak sector gives 
 an interesting  possibility to use it simultaneously in the role of the QCD axion, provided 
 gravitational contribution is absent due to additional chiral symmetries, or at least is reduced to 
 an acceptable level.  



 \section{discussions} 
 
 The three-form formulation allows for a simple physical interpretation of  the strong CP problem 
 and its solutions in terms of  Coulomb and Higgs phases of the QCD four-form electric field. 
 Any fundamental physics that jeopardizes this solution must take the three-form gauge theory out
 of the Higgs phase. This requires an additional massless three-form fields. 
 Gravity provides a natural candidate in form of the Chern-Simons three-form, but in the presence 
 of additional anomalous chiral symmetries, this form does not seem to be a suitable candidate 
 for unscreening the QCD three-form, since gravitational $\theta$-term can be rotated away. 
   String theory provides a variety of three-forms, but it also provides additional axions that Higgs these three-forms.  So at the end of the day the question comes down to the number of three-forms matching the 
   number of axions. 
  In the absence of additional three-forms gravity can still correct axion mass and couplings
  but these corrections are CP conserving.

\vspace{0.5cm}   

{\bf Acknowledgments}
\vspace{0.1cm} \\
 It is pleasure to thank Gregory Gabadadze, Michele Redi and  Neal Weiner for valuable discussions and comments.  This work is supported in part  by David and Lucile  Packard Foundation Fellowship for  Science and Engineering, and by NSF grant  PHY-0245068

\section{Appendix}

 In this appendix we shall give a more detailed discussion of why, in the absence of massless poles, 
 the axion potential is fully determined by (\ref{connection}).

In strongly coupled theories,  the classical actions (\ref{CBaction}) and (\ref{aFaction}) will receive quantum corrections.
These corrections will be represented  by high-dimensional operators regulated by the scale of the
corresponding physics.  
In the $B_{\mu\nu}$ description these operators should  respect the gauge (\ref{shiftB})
symmetry. We wish to establish an exact rule determining the axion potential 
from the quantum corrected action.  We shall attempt this by analyzing the most general
action in the description of the $B_{\mu\nu}$-field, and then calculating the axion potential 
by dualizing this effective action.  The external source $J_{\alpha\beta\gamma}$  plays no role in this derivation and we shall set it to zero. The generalized version of (\ref{CBactiondual})
can be  an arbitrary function of the  gauge invariant four-form  
$F_{\mu\alpha\beta\gamma}$ and the gauge invariant three-form 
$A_{\alpha\beta\gamma} \, = \, 
(P_{\alpha\beta\gamma}  \, - \, C_{\alpha\beta\gamma})$, and their derivatives.  The only requirement we 
shall impose on this action (other than the usual consistency requirements, such as the absence of 
negative norm states) is that it does not create any additional massless poles in the propagators.
That is,  no long-range correlations must exist in the effective theory. Under this assumption, the only part of the action that determines the axion potential, is the part that depends only on $F$ and not on its derivatives. In the absence of the massless poles, the higher derivatives
of $F$ and $A$ and their cross-couplings only contribute to the terms depending on the axion derivatives,  whereas  the axion potential is fully determined by the functional dependence on $F_{\mu\alpha\beta\gamma}$. 
 
 Consider a most general gauge invariant  Lagrangian describing a three-form field in the Higgs 
 phase.  As said above,  this Lagrangian is some general function of a gauge-ivariant scalar $F$ and 
and a three-form $A_{\alpha\beta\gamma}$ and possibly their derivatives.  We dualize this theory by 
introducing a Lagrange multiplier, axion, so that the resulting action is 
 \begin{equation}
\label{AFeffective1}
 S \, = \int_{3+1} \,  L(F, A_{\alpha\beta\gamma})  \, + \  {\Lambda^2 \over 24}  \,a\,\epsilon^{\mu\alpha\beta\gamma} \partial_{[\mu}\, P_{\alpha\beta\gamma]},  
\end{equation}
Varying this action with respect to $C, P$ and $a$, we get the following effective equations
of motion relating $F$ and $A$ to the axion field
\begin{equation}
\label{varyf}
{\delta S \over \delta F} \, = \, 24^{-1}\, (\kappa\,  - \, a)
\end{equation}
and 
\begin{equation}
\label{varya}
  {\delta S\over \delta  A_{\alpha\beta\gamma}}\, = \, \Lambda^2\epsilon^{\mu\alpha\beta\gamma} \partial_{\mu}\, a\,  
\end{equation}
Thus axion acts as a source, but $F$ is sourced by $a$ whereas $A$ is sourced by axion derivatives. 
In the absence of massless poles, on any Lorenz-symmetry preserving background $A$ can only depend on axion derivatives in such a way that $A \, \rightarrow \, 0$ when $\partial_{\mu} a \rightarrow 0$. Treating  $a$ as an external source,  if we  change its value at $t=0$ 
by a step function 
$a(t) \, =\, a_0\theta(t)$,  the response  in $F$ at $t=\infty$ will be 
\begin{equation}
\label{responce}
F(t=\infty) \, = \, {\rm inv}{\bf K}'(a_0)\, +\, {\rm massive~oscillations}.
\end{equation}
Whereas $A$ will only respond by massive oscillations. 
From the form of the axion equation
 \begin{equation}
\label{varyaxion}
 {\delta S \over \delta a} \, = \, F  \, + \, \epsilon^{\mu\alpha\beta\gamma} \partial_{[\mu}\, A_{\alpha\beta\gamma]},  
\end{equation}
it is then clear that only the axion dependence of $F$ will contribute into the axion potential
through the function ${\bf K}(F)$.  

 To see why the assumption about the absence of the massless poles is essential, consider 
the following Lagrangian  
 \begin{equation}
\label{lexample}
L\, =\, 12\,  A_{\alpha\beta\gamma} {\bf O}(\partial^2)  A^{\alpha\beta\gamma} \, + \, {1 \over 48} F^2 
\end{equation}
where ${\bf O}(\partial^2)$ is some differential operator. Variation of the action then gives
\begin{equation}
\label{varyf1}
 F \, = \, -\, (a\, - \, \kappa)\Lambda^2
\end{equation}
and 
\begin{equation}
\label{varya1}
  A_{\alpha\beta\gamma} \, = \,{\Lambda^2 \over 144} \epsilon^{\mu\alpha\beta\gamma} {\bf O}^{-1}(\partial^2)\partial_{\mu}\, a\,  
\end{equation}
and the effective equation for the axion field bocomes
\begin{equation}
\label{eqaxiono}
(a\, + \, \kappa) \Lambda^2 \, + \, \Lambda^2 {\bf O}^{-1}(\partial^2)\partial^2\, a  \, = \, 0
\end{equation}
From this expression it is obvious that unless ${\bf O}^{-1}$ operator has a pole at zero momentum
square, its contribution will vanish at $a =$constant, and there will be no contribution into the axion mass.   On the other hand, in the presence of a massless pole, the axion potential will receive an extra contribution. For instance, for ${\bf O} = { \Lambda^2\partial^2 \over M^2 + \partial^2}$, the axion equation becomes 
\begin{equation}
\label{eqaxiono1}
(a\, - \, \kappa) (\Lambda^2 \, + \, M^2) \, +  \, \partial^2\, a  \, = \, 0,
\end{equation}
with the extra contribution to the mass equal to $M^2$.  The existence of a massless pole, can only be 
achieved by integrating out a massless field, which contradicts to our assumption.

\end{document}